\documentstyle[12pt]{article}
\newcommand{\gsim}{ \raisebox{-.5ex}{\mbox{$\,\stackrel{>}{\sim}\,$}} }
\newcommand{\lsim}{ \raisebox{-.5ex}{\mbox{$\,\stackrel{<}{\sim}$\,}} }
\begin{document}
\begin{center}
{\large\bf Big Entropy Fluctuations
           in Statistical Equilibrium:\newline
           The Fluctuation Law}\\[5mm]
                   Boris Chirikov and Oleg Zhirov\\
{\it Budker Institute of Nuclear Physics \\
        630090 Novosibirsk, Russia}\\[1mm]      
        chirikov @ inp.nsk.su\\
        zhirov @ inp.nsk.su\\[5mm]
\end{center}

\begin{abstract} 
The structure of very complicated irregular "microscopic" (local) entropy
fluctuations around
a big separated "macroscopic" (global) fluctuation in the statistical 
equilibrium was studied
in numerical experiments on a simple\\ 2--freedom strongly chaotic 
Hamiltonian model 
described by the modified Arnold cat map.
A comparison of transient nonequilibrium rise and relaxation process of the
big fluctuation out of the statistical equilibrium with a nonequilibrium
steady state in a model without statistical equilibrium is considered and
discussed with respect to the so-called Fluctuation Law (or "theorem")
introduced and intensively studied recently in the latter case.
A new transient fluctuation law was found on the basis of a simple 
semiempirical theory developed.
Preliminary results of numerical experiments on some fractal properties
of the "microscopic" fluctuations are presented and briefly discussed.

\end{abstract}

\vspace{5mm}


\newpage


\section{Introduction: fluctuations in dynamical systems\\ 
         with and without statistical equilibrium}

\hspace*{\parindent} In a previous paper \cite{1} we presented numerical and
theoretical results of our studies into the peculiar properties of rare 
big ("macroscopic", or global) fluctuations out of the statistical equilibrium
as different from, and even opposite in a sense to, 
those of small stationary ("microscopic", or local) ones.
Particularly, the former are perfectly regular, on the average, 
symmetric in time with respect
to the fluctuation maximum, and described by simple kinetic
equations rather than by a sheer probability
of irregular "noise". Moreover, such fluctuations 
are not only perfectly regular by themselves but also
surprisingly stable against any perturbation, both regular and chaotic,
whatever its size. At first glance, it looks very strange
in a chaotic, highly unstable, dynamics. The resolution of
this apparent paradox is in that the dynamical instability of motion
does affect the fluctuation instant of time only. 
As to the fluctuation evolution, it is determined by the kinetics whatever its 
mechanism,
from purely dynamical one 
to a completely noisy, or stochastic (for discussion see \cite{1,2}).

A fairly simple picture of big fluctuations in the systems with statistical
equilibrium
was the basis of old Boltzmann's fluctuation
hypothesis for our Universe. As is well understood by now such a
hypothesis is completely incompatible with the present structure of the
Universe as it would immediately imply the notorious "heat death" (see, e.g.,
\cite{3}).
The principal solution of this problem, unknown to Boltzmann, is quite
clear by now, namely, the "equilibriumfree" models are wanted.
Various classes of such models are intensively studied today.
Moreover, the celebrated cosmic microwave background tells us that 
our Universe was born already in the state of a heat death which, 
however, fortunately to us all became unstable
due to the well--known Jeans gravitational instability
\cite{4}. This resulted in developing of a rich variety of collective
processes, or synergetics, the term recently introduced or, 
better to say, put in use by Haken \cite{5}.
The most important peculiarity of such a collective instability is in that
the overall relaxation with  ever increasing total 
entropy 
is accompanied by an also increasing phase space {\it inhomogeneity} 
of the system, particularly in temperature.
In other words, the whole system as well as its local parts become 
more and more {\it nonequilibrium} 
(for general discussion see, e.g., \cite{6,7,8}).
We stress that all these inhomogeneous nonequilibrium structures are not
big fluctuations like in statistical equilibrium but 
rather the result of a
regular collective instability,
so that they are immediately formed under a certain condition.
Besides, they are typically dissipative structures in Prigogine's 
term \cite{9} due to exchange of energy and entropy with the
{\it infinite} environment. The latter is the most important feature of such
processes, and at the same time the main difficulty in studying the
dynamics of those
models both theoretically and in numerical experiments which are so much
simpler for the systems with statistical equilibrium.

In spite of the essential restrictions, the simple
models with statistical equilibrium allow us to better understand the mechanism and role of fluctuations in the
statistical physics. Particularly, a vague problem of the initial conditions, still the apparently confusing (to many) "freedom",  
can be removed in such models.  
Following our previous paper \cite{1} we take
a different approach to the problem: instead discussing the "true" initial
conditions and/or a "necessary" restriction of those we start our numerical
experiments at {\it arbitrary} initial conditions
(most likely corresponding to the statistical equilibrium), 
and do {\it observe} what the dynamics 
and statistics of fluctuations is like. 
 Notice, however, that such an approach 
can be directly
applied to the fluctuations in finite systems
with statistical
equilibrium only (for discussion see \cite{6,2}).
In such, and only such, systems infinitely many 
big fluctuations grow up spontaneously 
{\it independent} of the initial conditions of 
the motion. This is similar
to the well--known Poincar\'e recurrences (for
discussion see \cite{1}). 

Recently, a new class of dynamical models 
has been developed by Evans, Hoover, Morriss, Nos\'e, and others 
(see, e.g., \cite{10}).
Some researchers still hope that such brand--new 
models
will help to resolve the "paradox of irreversibility". A more serious reason
for studying these models is in that they allow for a fairly simple
inclusion in a few--freedom model the infinitely dimensional "thermostat",
or "heat bath". This greatly facilitates both numerical experiments as well as
the theoretical analysis. The price is the strict restriction of such models
to the nonequilibrium
{\it steady states} only. Moreover, any collective processes of interacting
particles are also excluded, just those responsible for the very existence
of the nonequilibrium steady states.
In a more complicated Nos\'e - Hoover version of those models
these severe restrictions can be partly, but not completely, lifted.
Whether it would be sufficient for the inclusion of collective processes
remains, to our knowledge, an open question.

On the other hand, a nonequilibrium {\it steady state} studied in the
new models
is but a little, characteristic though, piece of the chaotic collective 
processes. In \cite{2} it was conjectured that the regularities in the fluctuations
found in a {\it nonequilibrium} steady state
can be applied, at least
qualitatively, to a small part of a big fluctuation in a statistical
equilibrium on both sides of the maximum. This conjecture was one of the
motivations for the present studies. 
The result turned out to be more interesting than expected
(Section 5). The conjecture was partly confirmed, indeed, 
but a new
surprising peculiarity of fluctuations was found which appears
to be even more generic, and which has been accidentally missed
in \cite{2}.
 
\section{The model}

In the present studies we make use of the same model as in
\cite{1}. For reader's convenience we briefly repeat its
description. It is specified
by the Arnold cat map\\ (see \cite{11,12}):
$$ 
   \begin{array}{ll}
   \overline{p}\,=\,p\,+\,x & mod\ 1 \\
   \overline{x}\,=\,x\,+\,\overline{p} & mod\ 1 
   \end{array} 
   \eqno (2.1)
$$
which is a linear canonical map on a unit torus. It has no parameters, and is
chaotic and even ergodic. The rate of the local exponential instability,
the Lyapunov exponent $\lambda =\ln{(3/2+\sqrt{5}/2)}=0.96$, implies
a fast correlation decay (relaxation)
with characteristic time $t_r\sim 1/\lambda\approx 1$.
Throughout the paper $t$ denotes the time in map's iterations.

As in \cite{1} we use a minor modification of this map:
$$ 
   \begin{array}{ll}
   \overline{p}\,=\,p\,+\,x\,-\,1/2 & mod\ C \\
   \overline{x}\,=\,x\,+\,\overline{p}\,-\,C/2 & mod\ 1 
   \end{array} 
   \eqno (2.2)
$$
where $C\gg 1$ is a circumference of the phase space torus. This allows us to study big fluctuations with diffusive 
rise/relaxation kinetics
in $p$ and characteristic time 
$t_p\sim C^2/4D_p\gg 1$ where $D_p=1/12$ is 
the diffusion rate. The relaxation time in $x$ does not depend on $C$ and remains short ($t_r\sim 1$)
so that subsequent values of $x$ are nearly uncorrelated.
This allows to restrict the statistical properies of the
motion to the action $p$ only. For example, the distribution
function becomes
$$
   f(x,p)\,\to\,f(p)\,=\,\int_0^1\,f(x,p)\,dx
   \eqno (2.3)
$$

The size of a big fluctuation in p is characterized by the
standard deviation $\sigma (t)$ for a group of $N$ trajectories (or noninteracting particles).

Below, as in \cite{1}, we shall consider a particular case of 
big fluctuations, namely one with the 
prescribed position in the phase space:
$$
   x_{fl}\,=\,x_0\,=\,{1\over 2}\,, \qquad  p_{fl}\,=\,p_0\,=\,{C\over 2}
   \eqno (2.4)
$$
at the (unstable) fixed point $x_0=1/2\,,\ p_0=C/2$
of map (2.2).
Then, the variance $v$ of the size in $p$
is determined by the relation
$$
   v\,=\,\sigma^2\,=\,\langle p^2\rangle\,-\,p_0^2 \eqno (2.5)
$$
where brackets $\langle ...\rangle $ denote the averaging over
$N$ trajectories.
In ergodic motion at equilibrium $v=v_{E}=C^2/12$. In what follows we will
use the dimensionless measure $\tilde{v}=v/v_{E}\to v$,
and omit tilde.

The variable $v(t)$ is especially convenient in diffusive approximation of the kinetic
equation  as it
is varying in proportion to time. Yet, 
its relation to the fundamental conception
of entropy is also important. The standard definition of the entropy,
which can be traced back to Boltzmann, reads in our case:
$$
   S(t)\,=\,-\,\langle\,\ln{f(p)}\,\rangle\,+S_0\,  
   \approx\,{1\over 2}\ln{v(t)}
   \eqno (2.6)
$$
where $f(p)$ is a coarse--grained distribution function, $S_0$ stands for an arbitrary constant, and the latter relation
is a very simple and convenient approximation found in \cite{1}
under the condition that the distribution $f(p)$ is the standard Gauss law (see Section 3 below):
$$
   f(p)\,=\,\frac{\exp{(-(p-p_0)^2/2v)}}{\sqrt{2\pi v}} 
   \eqno (2.7)
$$ 
and the constant $S_0$ is set to:
$$
   S_0\,=\,-\,{1\over 2}\ln{(2\pi e)}\,\approx\,-1.4189\,
   \approx\,-\sqrt{2}
   \eqno (2.8)
$$
Approximation in (2.6) holds on the most part of the big
fluctuation except a relatively
small domain near the equilibrium where the distribution in $p$
approaches the homogeneous one. The exact entropy 
(with constant (2.8)) in equilibrium is
$$
   S_{E}\,=\,-\,{1\over 2}\ln{\left({\pi e\over 6}\right)}\,
   \approx\,-0.18 \eqno (2.9)
$$
instead of zero in approximation (2.6). The difference is
relatively small, the smaller the bigger is the fluctuation.
In the main part of big fluctuation our simple relation for the entropy
reproduces the exact one to a surprising
accuracy (see Fig.4 in \cite{1}).

Notice that the distribution calculated from any {\it finite}
number of trajectories is always 
a coarse--grained one.
However, the direct application of the exact relation in
Eq.(2.6) requires too many trajectories,
especially for the fluctuation of a small size.
A great advantage of our aproximation is
in that the computation of $S$ does not require very many trajectories
as does the distribution function. In fact, even a single trajectory is
sufficient as Fig.1 in \cite{1} demonstrates !

A finite number of trajectories used for calculating the variance $v$
is a sort of the coarse--grained distribution,
as required in relation (2.6), but with a free bin size which can be 
arbitrarily small.
 
Now we can turn to the main subject of this paper,
the so--called fluctuation law (or Fluctuation "theorem").
We begin with the fundamental statistical property of the
dynamical model, the distribution function in $p$.
 
\section{The fluctuation law: $p$--distribution}
The "Fluctuation
Theorem" has been first obtained by Evans, Cohen and Morriss 
\cite{13} for a particular example of the nonequilibrium steady
state, using both the theory as well as numerics. For our
purposes it can be represented in the form:
$$
   \ln{\left({f(\Delta S)\over f(-\Delta S)}\right)}=
   \eta\cdot\Delta S\,, \qquad 
   \eta = {2\langle\Delta S\rangle\over \sigma^2}
   \eqno(3.1)   
$$
Here $f(\Delta S)$ is the probability of entropy 
(or entropy--like quantity as in \cite{13}) 
change $\Delta S$ in the ensemble
of trajectory segments of a fixed (appropriately scaled) duration $t_s$ 
for the
mean change $\langle\Delta S\rangle\ >\ 0$ and variance
$\sigma^2$, and the fluctuation parameter $\eta=1$ usually
taken to be unity.

By itself, the relation (3.1) is but a specific reduced
representation of
the normal probabilistic law, the Gaussian distribution,
in a suitable random variable $\Delta S$:
$$
   f(\Delta S) = {1\over \sqrt{2\pi\sigma^2}}\cdot
   \exp{\left(-{(\Delta S - \langle\Delta S\rangle )^2\over
   2\sigma^2}\right)} \eqno (3.2)
$$
shifted with respect to $\Delta S = 0$ due to the permanent entropy 
production at a constant rate in the
nonequilibrium steady state. The fluctuation law (3.1) immediately follows
from the normal law (3.2) but not vice versa.
However, the surprise (to many) was in that the probability
of {\it negative} ("abnormal" or "wrong") entropy change $\Delta S < 0$ 
(without time reversal!)
is generally not small at all reaching 50\% for sufficiently short $t_s$.
That is every second change may be "wrong" !?

Implicitly, all that is contained in the well developed
statistical theory (see, e.g., \cite{14}, Section 20, 
and \cite{15}). 
Nevertheless, the first
direct observation of this phenomenon in a nonequilibrium 
steady state \cite{13} has so much impressed
the authors that they even entitled the paper "Probability of
Second Law violations in shearing steady state". In fact, this is simply 
a sort of peculiar fluctuations discussed in
\cite{2}. 

In our opinion, the main
lesson one should learn from the Fluctuation law is that the entropy
evolution is generally {\it nonmonotonic} contrary to a common belief, 
still now.
The origin of this confusion is, perhaps,
in traditional conception of the fluctuations as a 
charcteristic on the microscopic scale well separated from
a much larger macroscopic scale with its averaged quantities
like the entropy production rate, for example.

In equilibrium steady state the macroscopic scale with the mean rate
$\langle\Delta S\rangle =0$ traditionally seems to be irrelevant
with the entropy trivially conserved. However,
in the nonequilibrium
steady state the macroscopic scale is represented by a finite
rate $\langle\Delta S\rangle\ >\ 0$, yet the "microscopic"
scale of the fluctuations can be well comparable with, and even
exceed, the former.

The border of nonmonotonic behavior is at $\Delta S = 0$
(no entropy rise at all) which corresponds to the probability
of "wrong" entropy changes $\Delta S < 0$
$$
   P_{wr}(F) = \int_{-\infty}^0\,f(s,F)\,ds  \eqno(3.3)
$$
Here $F=\langle\Delta S\rangle /\sigma$ is a new parameter for the "right/wrong" crossover
in the entropy variation sign at $|F|=F_{cro}\sim 1$ when
the probability $P_{wr}$ is large.

If a finite--dimensional Hamiltonian system admits the (stable)
statistical equilibrium as in our model (2.2) here the
overall ($t\to\infty$) average entropy rate 
$\langle\Delta S\rangle\ = 0$ for any $t_s$. However, on a
finite time scale $t_p\sim C^2$ (see Section 2) of a nonequilibrium 
relaxation to the
equilibrium the local $\langle\Delta S\rangle\ >\ 0$ as in the
nonequilibrium steady state, but temporally. On this time scale
the fluctuations were conjectured \cite{2} to obey the law
similar to the nonequilibrium one provided $t_s\ll t_p$.

The entropy--like quantity in our problem is variance $v$, Eq.(2.5),
which monotonically depends on the entropy (2.6).
The time dependence $v(t)$ in a big fluctuation was computed as follows
(for details see \cite{1,2}). 
The data were obtained from simultaneous running of $N$ trajectories for 
very
long time in order to collect sufficiently many fluctuations for the reliable 
separation of
the regular part of the fluctuation, or the kinetic subdynamics 
in Balescu's term
(see \cite{16} and references therein), from the stationary fluctuations,
the main subject of the present studies (see below).
The separation was done by the plain
averaging of individual $v_i(t)$ values ($i=1,...,n$) over all $n$ 
fluctuations collected in a run. 
The size of the fluctuations was
fixed by the condition that current
$$
   v(t) < v_b \eqno (3.4)
$$
at some time instant $t\approx t_i$, the moment of a fluctuation. 
This condition
determines, in fact, the border of the whole fluctuation
domain: $0 < v < v_b$.
The event of entering this domain is the macroscopic "cause" of the
fluctuation whose obvious
"effect" {\it will be subsequent} relaxation to the equilibrium.
However, and this is the main point of our philosophy, the second "effect"
of the same "cause" {\it was preceding} rise of the fluctuation in apparent 
contradiction
with the "causality principle" (for discussion see \cite{2}).
In any event, the second effect requires the permanent memory 
of trajectories within some time window $w$.
Typically,
$w\gsim C^2$, the total diffusion time, was chosen (see Section 2).
After fixing the current $t_i$ value the computation within
the same window $w$ had been continued, and only then
the search for the next fluctuation was resumed.

Two examples of big global fluctuations are shown in Fig.1. They differ by the
number of separate fluctuations in each run for averaging.
While for a larger $n=1137$ the dependence 
$\langle v(\tau )\rangle$
looks rather smooth, in the second
run, with $n=32$ only, the stationary local fluctuations around the averaged 
global one are clearly seen.

The average anti--diffusive/diffusive kinetics from/to the statistical
equilibrium (horizontal straight line) is shown by the two
wiggly curves.
A smooth solid line is semiempirical relation (3.13) to be discussed below
together with the expected kinetics near the maximum, Eq.(3.14),
(two oblique straight lines). The mean variance
$\langle v(t-t_i)\rangle$ is doubly averaged in both the number of trajectories
$N$ (see Eq.(2.5)) and that of recurrent fluctuations $n$.

In this Section we consider the statistics of the original dynamical variable, 
the dimensionless action 
$\tilde{p}=2\sqrt{3}p/C\to p$ (see Section 2),
with respect to the fixed fluctuation position $p_0=\sqrt{3}$
, or of the quantity $u=p-p_0$. Its two first moments, in the
limit $N\cdot n\to\infty$, are
$$
   \langle u(\tau )\rangle\,=\,0\,, \qquad \langle u^2(\tau )\rangle\,
   =\,\langle v(\tau )\rangle \eqno (3.5)   
$$
Neglecting all dynamical correlations (see Section 2) would imply the standard
Gaussian distribution
$$
   G(u)\,=\,\frac{\exp{(-u^2/2v(\tau ))}}{\sqrt{2\pi v(\tau )}}\,=\,
            \frac{\exp{(-g^2/2)}}{\sqrt{2\pi}}
   \eqno (3.6)   
$$
(cf. Eq.(2.7)) provided a free (unbounded) anti--diffusion/diffusion. Moreover,
in a new, Gaussian, variable $g=u/\sqrt{v}$ the distribution would
not depend on time either.
This allows for a considerable increasing of the statistics in computation
of the actual distribution $f(u)$ by summing up the data in a certain interval
of $\langle v(\tau )\rangle$ shown in Fig.1.

The result is presented in Fig.2
in the form of the ratio $f(u)/G(u)$ as a function of the Gaussian variable
$g^2=u^2/v$. Within a moderate $g^2\lsim 5$ the ratio is close to unity
as expected for a free $x$--uncorrelated diffusion.
However, for larger deviations the ratio is progressively decreasing while
the shape of distribution $f(u)$ remains Gaussian within this region.
Qualitatively, it is similar to the distortion observed in a nonequilibrium
steady state (see \cite{2}, Figs.2 and 3) but of the opposite sign.
Moreover, the crossover between the two regions
$$
   g^2_{cro}\,\approx\,4.5 \eqno (3.7)
$$
is nearly the same in both processes, and also for the both examples in Fig.2 here.
In \cite{2} it was conjectured that the origin of such a distortion
might be some peculiar effect of dynamical correlations in $x$ which are small
but nonzero (Section 2). Yet, this still remains an open question.

In addition, the second crossover does appear which is not as "universal"
as the first one but also demonstrates the local Gaussian shape of the
empirical distributions that is a straight line in the
semi--log scale. The first crossover does not depend on the
averaging domain but the slope of the distribution above the
crossover (in the second region) does so. To the contrary,
the second crossover depends on the averaging conditions but
the slope in the third region does not. 
Apparently, the third region represents the effect of the
boundary condition on the long tails of the distribution.
All these interesting
peculiarities of a very specific "triple" Gaussian distribution
require further studies.

In the first region $g^2\lsim 5$, which comprises approximately
97\% of the total probability, the distribution is reasonably
close to the Gaussian one, and this explains a surprising
accuracy of simple relation (2.6) for the entropy (see \cite{2},
Fig.6). 

Near the equilibrium the distribution can be calculated from
a simple diffusion equation:
$$
   \frac{\partial f(\tau ,u)}{\partial t}\,=\,{D\over 2}\cdot
   \frac{\partial^2 f(\tau ,u)}{\partial u^2} \eqno (3.8)
$$
where $D=1/C^2$ is the diffusion rate.
For periodic boundary conditions\\ $f(\tau ,u+2\sqrt{3})=
f(\tau ,u)$, used in numerical experiments with model (2.2),
and for the fluctuation symmetry with respect to $u=0$ (see \cite{1}) the eigenfunctions and eigenvalues of Eq.(3.8) are
respectively:
$$
   \phi_m\,=\,\cos{\left(\frac{\pi mu}{\sqrt{3}}\right)}\,,
   \qquad \gamma_m\,=\,{D\over 2}\cdot{\pi^2m^2\over 3}\,=\,
   {\pi^2m^2\over 6\,C^2} \eqno (3.9)
$$
where $m=0,1,...$ is any integer. 

In the first approximation, we can assume the initial
distribution  $f(0,u)\,=\,\delta (u)$ to be a
$\delta$--function. Then, the solution of diffusion equation
(3.8) on both sides of the fluctuation maximum at $\tau =0$
is represented by a series:
$$
   f(|\tau |,u)\,=\,{1\over 2\sqrt{3}}\,+\,{1\over \sqrt{3}}
   \sum_{m=1}^{\infty} {\rm e}^{-\gamma_m |\tau |}
   \cos{\left(\frac{\pi mu}{\sqrt{3}}\right)}
   \eqno (3.10)
$$
Here the first term describes the ergodic equilibrium, and
the others do so for the rising/decay of a big fluctuation,
on the average.

Moreover, we could further
approximate the initial $\delta$--function 
by a Gaussian distribution
$$
   \delta (u)\,\approx\,\frac{\exp{(-u^2/2v_b)}}{\sqrt{2\pi v_b}}
   \eqno (3.11) 
$$
where $v_b\ll 1$ is the size of the fluctuation domain (3.4).
Then, one would expect the Gaussian distribution to
persist up to $|\tau |\sim 1/\gamma_1\sim C^2$ when lower modes
of the solution (3.10) come into play. In fact, according to
our numerical experiments the Gaussian distribution is
considerably distorted also for small $|\tau |\lsim 50$.
Apparently, this distortion is caused by the
selection condition (3.4) which cut out large distribution
fluctuations.

Near the equilibrium the solution provides a simple relation
for the main quantity in the problem, the variance
$$
   v(|\tau |)\,=\,\langle u^2\rangle\,\approx\,
   1\,-\,{12\over \pi^2}\cdot
   \exp{\left(-{\pi^2|\tau |\over 6\,C^2}\right)}
    \eqno (3.12)
$$
which is close to a semiempirical relation
in Fig.1:
$$
   v_f(|\tau |)\,\approx\,
   1\,-\,\exp{\left(-{|\tau |\over 0.65\,C^2}\right)}
   \eqno (3.13)
$$
Particularly, the most important factor in the exponential
(3.12) $6/\pi^2=0.608$ is only 6\% less than the empirical one
$0.65$ in (3.13). 
On the other hand, both expressions, Eqs.(3.12) and (3.13),
are also close to a simple relation $v(|\tau |)\,\approx\,|\tau |/C^2$,
directly derived from the diffusion rate $D=1/C^2$,
upon a small correction
$$
   v_f(|\tau |)\,\approx\,|\tau |/C^2\,+\,0.05\,, \qquad
   |\tau |\,\lsim w/2 
   \eqno (3.14)
$$
introduced in \cite{1} to describe the
effect of a finite fluctuation size at $\tau =0$ determined
by the selection condition (3.4).

In conclusion of this Section we stress again that in spite of
considerable deviations on the tails the distribution $f(u)$
remains close to the normal (Gauss) law in its main part
comprising 97\% of the total probability. This will be 
essential in the next Section.

Now we can turn to the main problem of the present studies,
the fluctuations of our entropy--like quantity, the
variance $v(\tau )$.

\section{The fluctuation law:\\
         "Wrong" macroscopic entropy variations\\
         of both signs}
The so--called "Fluctuation Law", recently introduced in the
studies of the nonequilibrium steady state, was discused at
the beginning of previous Section. In our problem it can be
related to an entropy--like quantity, the variance 
$v(|\tau |)$. So, first of all, we
consider here the statistics of this variable.

\subsection{$\chi^2$--distribution}
Under condition of the normal distribution in $u$, which is
a good approximation (see previous Section), the statistical
properties of the sum of $u^2$ values are described by the
so--called $\chi^2$--distribution (see, e.g., \cite{17}):
$$
   \chi (s)\,=\,\frac{s^m\,{\rm e}^{-s}}
   {\Gamma (m+1)}\,\to\,\chi (m)\cdot\exp{[-(s-m)^2/2m]}
   \eqno (4.1)
$$
where  random quantity
$$
   s\,=\,{1\over 2}\sum_{i=1}^k u_i^2\,=\,
   {k\tilde{v}\over 2} \eqno(4.2)
$$
$\Gamma (x)$ is the gamma--function,
$m=k/2-1$, and variance 
$\tilde{v}=v/\langle v(|\tau |)\rangle\to v$ is the random
variable now normalized to its mean according to the Gaussian
distribution in $p$ (Section 3). Again, we omit tilde in what
follows (cf. Section 2).
The first expression in (4.1) is the exact distribution with
three main characteristics:
$$
   s_{max}\,=\,m\,, \qquad  \langle s\rangle\,=\,m+1\,,
   \qquad  v_s\,=\,\langle s^2\rangle\,-\,\langle s\rangle^2\,
   =\,{2\over k}\,\langle s\rangle^2\,=\,
   {k\over 2}
   \eqno (4.3)
$$
the maximum of probability density, mean $s$, and its
variance.
One should not confuse the latter, which is a characteristic
of the random variable $s$, with another variance $\tilde{v}
\to v$ described above.
The latter expression in (4.1) is a Gaussian approximation
to the former for large number $k\gg 1$ of terms in sum (4.2).
Notice the shift $\langle s\rangle - s_{max} = 1\ll k$.
The approximation was chosen in such a way to provide the best
Gaussian description for the distribution cap that is most
important in our studies below.
To this end,
both the position and height of the top as well as the second
derivative over there were fixed to be exactly equal in the
two distributions. Interestingly, the two normalizing factors,
in Eq.(4.1) and the standard Gaussian one, are very close
$$
   \chi (m)\cdot\sqrt{2\pi m}\,=\,1\,-\,{1\over 12m}\,+\,...
   \eqno (4.4)
$$
even for small $m$.

In Fig.3 a comparison of the two distributions for $k=10$ is 
shown. The very cap looks perfect but the progressive deviation
below is primarily due to asymmetry of the $\chi^2$--distribution. In what follows we restrict ourselves
to this optimized Gaussian approximation.

Now, we need to transform the distribution to the
main dynamical (and chaotic) variable in our problem,
the random variance $v=2s/k$, Eq.(4.2). We obtain:
$$
   \chi (v)\,\approx\,\frac{\exp{[-(v-v_0)^2/2\sigma_v^2]}}
   {\sqrt{2\pi \sigma_v^2}}
   \eqno (4.5)   
$$
where
$$
   v_0\,=\,{m\over m\,+\,1}\,=\,1\,-\,{2\over k}\,, \qquad
   \sigma_v^2\,=\,{m\over (m\,+\,1)^2}\,
   \approx\,{2\over k}
   \eqno (4.6)
$$
the distribution maximum, and $v$ variance, respectively.
Again, notice a small shift 
$v_0-\langle v\rangle = -2/k \approx -\sigma_v^2$.

As was explained in Section 3 the variance $v(\tau )$ was
computed in numerical experiments by the two averagings.
First, in each realization of the big fluctuation the current
averaging is done over a group of $N\lsim 10$ trajectories. 
On this stage $k=N$ should be substituted in Eq.(4.6). 
Since technically it is very difficult to increase $N$ the
accuracy of our approximation (4.5) is generally rather poor
(see Fig.3).
Moreover, this trouble goes over to the second averaging of
$n$  successive
realizations of the big fluctuation, no matter how large is
the number $n$.
As different realizations are statisticaly independent,
formally the parameter $k=N\cdot n$ increases by $n$ times
but the problem is in actual deviation of the approximation
(4.5) from the true unknown distribution except, hopefully,
the distribution cap. In any event, all we can do at the
moment is aplying Eq.(4.5) with the latter value of
$k=Nn$. 

\subsection{The probability of "wrong" entropy variation}
In the nonequilibrium steady state the entropy is always
{\it increasing} on the average, and the "wrong" variation means
sporadic {\it decreases} of the entropy on some finite
time intervals (see, e.g., \cite{10,2} and references therein).
In our problem of a big finite fluctuation out of statistical
equilibrium the situation is more interesting. Namely,
the entropy is both {\it decreasing} and {\it increasing},
on the average, during the corresponding anti--diffusion and
diffusion stage of the fluctuation, and respectively the "wrong"
entropy variation is both its temporary
{\it increase} or {\it decrease}.

Here we present some preliminary results of numerical experiments on the local fluctuations around the big (global)
fluctuation as explained in Section 3.

The computation procedure of obtaining the data for the
fluctuation law was as follows. The whole time window $2w$
was subdivided in a number of segments of length $t_s$
(iterations), and a change $\Delta v(\tau )$ of the doubly
average variance (our entropy--like quantity) on each one was
calculated. To suppress still large local fluctuations a new
double averaging was applied. The first one was defined as
$$
   \overline{\Delta v(\tau_j)}\,=\,
   {1\over t_s}\sum_{\tau =\tau_j}^{\tau_j+t_s}
   \Delta v(\tau ) \eqno (4.7)
$$
Since successive $\Delta v(\tau )$ are not independent this
averaging turned out to be insufficient. The second one was
done over $L$ successive segments:
$$
   \overline{\overline{\Delta v(\tau_i)}}\,=\,
   {1\over L}\sum_{j=1}^L \overline{\Delta v(\tau_j)}
   \eqno (4.8)
$$
with the final result ascribed to $\tau_i = \tau_j + t_sL/2$,
the center of the full averaging interval.

Of two different formulation of the fluctuation law, Eqs.(3.2)
and (3.3), we have chosen the latter one for the integral 
probability as more reliable because of much less fluctuations.
For calculation of probability $P_{wr}$ of "wrong" 
$\Delta v(\tau )$ the number of segments satisfying the
condition 
$$
   \tau\cdot\Delta v(\tau )\,<\,0
   \eqno (4.9)
$$
was counted. 

An example of time dependence for such a probability is
presented in Fig.4 for segment length $t_s=30$ and averaging
moving window size $L=30$ using the data in Fig.1 with 
$n=32$ and $N=4$. The empirical results are shown by points
the number of which is $2w/t_s\approx 650$. 
For large $|\tau |\gsim 2000$ the probability $P_{wr}\approx 0.5$
oscillates around 50\% as expected in the equilibrium.
However, near the fluctuation maximum the probability of "wrong"
entropy changes rapidly drops almost down to zero. 
This is a result of the average (macroscopic) entropy
destruction/production in course of anti--diffusion/diffusion.

The upper solid curve in Fig.4 represents a simple theory described above (with $b=1$, see Eq.(4.10)).
It is based on Eq.(3.3) with the distribution
$f(s)\approx \chi (v)$ in the Gaussian approximation (4.5).
Then, the fluctuation parameter in Eq.(3.3) becomes:
$$
   F(\tau )\,=\,{|\langle\Delta v_f\rangle |\over \sigma_f}\,\approx\,
   \frac{t_s(1\,-\,v_f(\tau ))/aC^2}{v_f(\tau )
   \sqrt{2\sigma_v^2/b}} \eqno (4.10)
$$
Here the average change and the variance $\sigma_f^2$ are expressed via 
the semiempirical function $v_f(\tau )$ with parameter $a=0.65$,
Eq.(3.13), and
parameter $\sigma_v^2$ with $k=nN$ in Eq.(4.5) as follows:
$$
   \langle\Delta S\rangle\,\to\,
   |\langle\Delta v_f\rangle |\,\approx\,t_s{dv_f(\tau )\over d\tau}\,\approx\,
   {t_s(1\,-\,v_f(\tau ))\over aC^2}\,, \quad
   \sigma_f^2\,\approx\,2\cdot{\sigma_v^2\over b}\cdot
   v_f^2\,=\,{4v_f^2\over nNb}
   \eqno (4.11)
$$
In the latter relation the factor 2 is introduced assuming
statistically independent fluctuations on both ends of the
segment $t_s$. The calculation of theoretical probability
$P_{wr}$ was performed by the same double averaging, Eqs.(4.8),
of the integral (3.3) with parameter (4.10) as for the
empirical data. 

The exact function (3.3) is unknown, as was
explained above, but under assumption of Gaussian $v$
fluctuations it becomes the standard error integral which
can be explicitly calculated. Moreover,
we managed to find a very
simple and surprisingly accurate approximation for the error
integral which in a particular case of Eq.(3.3) takes the form:
$$
   P_{wr}(F)\,\approx\,\frac{\exp{(-F^2/2)}}{2(F\,+\,1)}
   \eqno (4.12)
$$
A comparison with the exact integral is shown in Fig.5.
The relative accuracy $|\Delta P/P|<0.05$ is better than 5\%
in a huge range of $P\gsim 10^{-4}$ larger than 5 orders of 
magnitude! Asymptotically, as $F\to\infty$ the errror
$|\Delta P/P|\to \sqrt{\pi /2} - 1\approx 0.25$ increases
but still remains surprisingly small.

Coming back to the comparison of theory (4.12) with the
empirical data in Fig.4 we see that the agreement
is rather poor.
Qualitatively, the theory describes the phenomenon but 
considerably overestimates the probability of the "wrong"
entropy changes. This is why we have had to introduce an
empirical factor $b$ into our simple theory (4.10). 
To achieve the agreement with the empirical data we have to
take the value of this factor as large as $b=30$ 
(the lower solid curve in Fig.4).
It hardly could be explained by a distortion of the Gaussian
distribution we neglected.
To understand the origin of such a discrepancy we have
undertaken a study of various statistical properties of
the $v$ fluctuations.

\subsection{Various $v$--statistics}
We begin with the so--called {\it current} fluctuations
that is the fluctuations at a fixed $\tau$. These fluctuations
are approximately described by the optimized Gaussian distribution (4.5) in dimensionless variable 
$v/\langle v(\tau )\rangle\approx v/v_f(\tau )$ where $v_f$
is given by the approximate relation (3.13).
The first two moments of the distribution were computed, namely
the expected mean shift specific for the optimized Gaussian distribution
$$
   \langle{v\over v_f}\,-\,1\rangle\,\approx\,\sigma_v^2\,\pm\,\sigma_v
   \eqno (4.13)
$$
and the variance
$$
   \langle\sigma_v^2\rangle\,
   \approx\,{2\over k}\,=\,{2\over nN} \eqno (4.14)
$$
which also determines the standard deviation for the shift in
Eq.(4.13). The empirical specific shift turned out to be,
indeed, very close to theoretical one, the ratio of both
being 1.05.
However, the corresponding ratio 0.36 for the variance itself
indicates a considerable disagreement with the theory.
To further elucidate this point we have computed the time
dependence for both characteristics using our standard procedure
of the double averaging (Section 4.2).
The results presented
in Fig.6 clearly demonstrate a regular increase of both the
shift and variance at small $|\tau |$, just in the region
which is most important in our studies (see Fig.4).
Partly, it is explained by a poor approximation (3.13) we use.
To cope with this difficulty we have even omitted the region
$|\tau |<200$ in averaging but this did not help in full.

In any event, the observed {\it three}fold decrease of the current variance cannot explain the {\it thirty}fold (!) decrease required
to agree the theory with the empirical data in Fig.4.
So, we turned to investigation of the main parameter of the theory $F(\tau )$, Eq.(4.10).
To this end, we have computed different {\it segment} fluctuations
in comparison with the theoretical prediction.
Specifically, we calculated three average ratios (see Eqs.(4.7) and
(4.8)):
$$
   [\ \overline{\overline{\Delta v}}\ ]_R\,=\,
   {1\over L}\sum \frac{\overline{\Delta v}}{\Delta v_f}
   \eqno (4.15a)   
$$
for the average change $\Delta v$ per segment,
$$
   [\ \overline{\overline{(\Delta v)^2}}\ ]_R\,=\,
   {1\over L}\sum {\overline{\sigma^2}\over \sigma_f^2}
   \eqno (4.15b)
$$
for the variance of $\Delta v$ per segment, and
$$
   [\ \overline{\overline{F^2}}\ ]_R\,=\,
   {1\over L}\sum
   \frac{\left(\overline{\Delta v}\right)^2}
   {\overline{\sigma^2}}
   \cdot\frac{\sigma_f^2}{(\Delta v_f)^2}
   \eqno (4.15c)
$$
for the ratio of the former that is for the fluctuation
parameter squared. The theoretical quantities with sub $f$ were
taken from Eq.(4.11) and applied to the second averaging only.


The results are presented in Fig.7 for three values of the
segment length $t_s=20,\,30,\,40$. The first ratio (4.15a)
for the mean $\Delta v$
is shown by the middle solid curves, and it seems 
in a reasonable agreement with the theory. On the contrary,
the variance (4.15b) (lower dashed curves)
is about two orders of magnitude less
than expected except the central region of small $|\tau |$
where it is much larger and strongly depends on $t_s$.
Particularly, for $t_s=30$ the variance is close to the
theoretical one in disagreement with the upper theoretical
curve in Fig.4. The difference between the two series of data is
in the averaging procedure. In Fig.7 it was a separate averaging
of the variance only while in Fig.4 the averaging included
the ratio of the mean change squared to the variance 
of $\Delta v$. The latter is chracterised by the third ratio
(4.15c) shown in Fig.7 by the upper solid curves.
This data do better describe the fluctuation parameter 
$F(\tau )$. Yet, the empirical parameter $b\gsim 100$ is now
at least three times as big compared with $b=30$ in Fig.4.
Apparently, the remaining discrepancy is explained by still
more complicated averaging of the probability $P_{wr}(F)$
as well as other approximations in the theory.

In spite of all these difficulties the theory developed
provides a consistent picture of the segment fluctuations
including their most intriguing part of the "wrong" entropy
changes. Moreover, an approximate scaling of the empirical
data can be inferred from Eq.(4.10)
with respect to the dimensionless variable $\tau /t_s$.
At least, this is possible in the range $t_s= 20\ -\ 40$
where the curves in all three groups in Fig.7 are close 
(besides small $|\tau |$ for the second ratio which is unimportant for the final conclusion).

The scaling is based on a simple approximation 
$v_f(\tau )\,\approx\,|\tau |/C^2$ (see Eq.(3.14)) which holds
for $v < 0.5$, or $|\tau | < C^2/2$. Then, Eqs.(4.10) and (4.11)
imply
$$
   F(\tau )\,\approx\,\frac{t_s/C^2}{(|\tau |/C^2)\cdot
   \sqrt{4/nNb}}\,=\,\frac{t_s}{|\tau |}\cdot 
   \sqrt{{nNb\over 4}}
   \eqno (4.16)
$$
This scaling is shown in Fig.8 for the interval 
$t_s = 10\ -\ 40$, and $b=30$ (cf. Fig.4).
In spite of some divergence of the theoretical curves for 
large $|\tau |/t_s \gsim C^2/2t_s$ the scaling describes the
probability $P_{wr}(\tau /t_s)$ fairly well within the
fluctuations, except perhaps the case for $t_s=10$.

\section{Conclusion: A new conjecture}
In the present paper we report the preliminary results of our
investigation into the local (segment) fluctuations in the
transient steady state supported by a big regular fluctuation
out of the statistical equilibrium (see Fig.1).
One of the
motivations for these studies was a conjecture \cite{2}
that the fluctuation properties, particularly the
fluctuation law, formulaed and intensively investigated in
the nonequilibrium steady state, are similar to those in the
transient steady state of a dynamical system with statistical
equilibrium.
Our original goal was either to confirm or to disprove this
conjecture. A preliminary answer to this question, we have
reached so far, turns out to be more interesting than just
the plain yes or no. 

In the nonequilibrium steady state the entropy is always
increasing on the average, and the "wrong" variation means
sporadic decreases of the entropy on some finite
time intervals (see, e.g., \cite{10,2} and references therein).
In our problem 
the entropy is not only increasing, 
on the average, during the corresponding diffusion 
stage of a big fluctuation but also decreasing on the
previous stage of anti--diffusion, so that the notions 
"right" and "wrong" exchange upon crossing the fluctuation
maximum.

We did observe, indeed, such a generalized fluctuation law
(in our formulation (3.3), see Figs.4 and 8). Moreover, 
we have developed a simple theory, Eqs.(4.12) and (4.10),
which qualiatively describes this law.
However, to reach a quantitative agreement with the empirical data we had to introduce into the theory a surprisingly big 
fitting parameter $b=30$ instead of expected $b=1$.

Investigation into this difficulty revealed that the problem
is in unexpectedly small variance of the segment fluctuations
of the entropy--like change $\Delta v$ (see Fig.7).
Thus, the main physical question to be answered is
the origin and mechanism of such a strong suppression of the
segment fluctuations.

A possible answer to this question, suggested by a careful
inspection of the fluctuation structure in Fig.1, is the
following. 
Indeed, the wiggly curve in this figure demonstrates
some very complicated fractal structure of the local
fluctuations. For the problem in question,
the most important feature of this structure
is a large variety of its time
scales up to that of the underlying big fluctuation itself.
This is especially clear from the picture of the maximal local
fluctuations presented in Fig.9 for two separated realizations
of a big fluctuation. Here, without averaging over many
realizations, it is even difficult to discern the local 
fluctuations from global ones,
particularly by their shape. Notice that both
can be not only negative with respect to the equilibrium but
also positive, up to $v(\tau )=3$ (see Section 2). In the
latter extreme case all the trajectories are concentrated near
$p=0\ mod\ 1$ that is the position of such fluctuations differs
from one for $\langle v(\tau )\rangle\ll 1$ studied in this
paper. This would require redifinition of the variance $v$
(2.5), and of approximate relation (2.6) for the entropy.
In any case, the entropy of a big fluctuation does first
decrease, and only then grows back to the equilibrium,
contrary to an immediate impression from Fig.9.

Essentially, the mechanism of large--scale local fluctuations
is the same as (or, at least, very similar to) that for the
big fluctuations that is in both cases it is the 
anti--diffusion/diffusion
with roughly the same rate.
This would imply the strong correlations between the values
of $v(\tau )$ for {\it different} $\tau$
in spite of statistically independent realizations of $v(\tau )$
for any {\it fixed} $\tau$.
Apparently, these very correlations is the ultimate origin of
considerable suppression
of the segment fluctuations.

An example of this correlation is presented in Fig.10.
It describes the difference
$$
   v_c\,=\,v(\tau )\,-\,\langle v(\tau )\rangle\,\approx\,
   v(\tau )\,-\,v_f(\tau ) \eqno (5.1)
$$
where $v_f$ is aproximation (3.13). Since the latter is rather
poor for small $\tau$ the calculation of correlation was
performed starting at $\tau_0=300$ in a window 
$T=2^{13}=8192$ which is the period of $v_c(\tau )$ for
subsequent Fourier transforms.
The correlation obtained in such a way is well described by a simple
empirical relation
$$
   K(t)\,\approx\,K_f(t)\,=\,
   A\cdot\exp{(-t/t_K)}\,+\,R \eqno (5.2)
$$
where $t=|\tau -\tau'|,\ A=0.9,\ R=0.1$, 
and the characteristic time of 
approximately exponential decay $t_K=250$.

Within the range of the segment length $t_s\leq 40$ in Fig.8
the correlation decay is less than 20\%, and this explains
the observed approximate scaling. However, the origin of a long residual correlation $R$, comprising both the free diffusion
as well as the equilibrium (cf. Fig.1), remains unclear.

Certainly, a more complete mathematical analysis and physical
interpretation are still to be done.

Another interesting question is why the local fluctuations
in a nonequilibrium steady state do not show any suppression
(see, e.g., Fig.2 in \cite{2}). By eye, the time dependence of
the entropy there (Fig.4 in \cite{2}) looks like a fractal
one similar to that in Fig.1 above. What is the difference?

We conjecture that the main origin of an essentially different
fractal structure in \cite{2} is in a "minor" modification
of the dynamical model there. Namely, like in the present paper
and in previous publication \cite{1} the model included parameter $C$ 
but just for the main study of the nonequilibrium
steady state it was set to one ($C=1$) without paying much
attention to this particular case. 
As a result, the relaxation of local fluctuations in $p$
became ballistic (fast) as in $x$, and hence the 
{\it correlations} were suppressed which ensured the standard
Gaussian statistics of the fluctuations.

For the problem under consideration here this choice would be
very difficult in case of the diffusive kinetics of the global
(big) fluctuations. But then, with $C\gg 1$, the local
fluctuations become also diffusive (see Figs. 1 and 9), hence
strong correlations and suppression of the latter
resulting, particularly, in a considerable decrease of the
probability of "wrong" entropy variation.

We conjecture that for large parameter $C\gg 1$ in the model
for nonequilibrium steady state the statistics of the local
fluctuations, particularly the fluctuation law, will be much
more interesting for investigation, and more difficult too.
Certainly, the whole problem deserves farther studies.

\newpage   

\begin{center}  Figure captions   \end{center}

\begin{itemize}

\item[Fig.1] 
Two big fluctuations averaged over different numbers
of repetitions (recurrences) in each run (see text):
wiggly lines show the time dependence 
of the mean variance
$\langle v(t-t_i)\rangle$ around the fluctuation maxima; 
a smooth solid line is semiempirical relation, Eq.(3.13), for the 
anti--diffusion/diffusion kinetics of fluctuation;
two oblique straight lines represent
the expected diffusive kinetics near the maximum, Eq.(3.14), and the
horizontal straight
line is the equilibrium.
Run parameters and results are respectively: $C=50,\ N=5/4,\ 
v_b=0.0256\ /\ 0.0034,\ n=1137\ /\ 32,\ w=10000$.
Average period between successive fluctuations 
$\langle P\rangle\approx 7.7\times 10^5\ /\  2.3\times 10^6$ iterations.
Two short dotted lines at the bottom indicate the range of the measurement
of $p$--distribution in Fig.2.

\item[Fig.2]
A triple "Gaussian" distribution for a big global fluctuation.
The ratio of empirical to the standard Gaussian distribution vs. the
Gaussian variable $g^2=u^2/v$ is shown for
two averaging domains (see Fig.1): $\tau = 50 - 350$
(wiggly line) and $\tau = 50 - 100$ (dots). 
Oblique straight lines demonstrate the Gaussian shape in
all three regions of the empirical distributions. 
Run parameters and results are respectively: $C=50,\ N=5,\ 
v_b=0.0256,\ n=32598,\ w=10000,\ 
\langle P\rangle\approx 7.7\times 10^5$; averaging bin size $\Delta u=0.05$.

\item[Fig.3]
Comparison of $\chi^2$--distribution (solid line)
and its best Gaussian approximation (dotted line),
Eq.(4.1): $k=10\ (m=4)$.

\item[Fig.4] 
Probability of "wrong" entropy change in ensemble of
trajectory segments for a big fluctuation in Fig.1:
$n=32,\ t_s=L=30$; points are empirical data; upper solid curve is
theory, Eq.(4.12), with empirical parameter in Eq.(4.10) $b=1$;
lower curve is the same for $b=30$.

\item[Fig.5]
Comparison of the exact integral (3.3) for the Gaussian
distribution $f(s)$ (solid curve) with approximation (4.12)
(circles); dashed line is relative accuracy of the approximation.

\item[Fig.6]
Empirical time dependence of the shift, Eq.(4.13) (solid line),
and that of the standard deviation 
(two dashed curves) 
for the data in Fig.1 with $n=32,\ t_s=L=30$.
The dashed horizontal straight line is the
mean reduced shift
$\langle (v/v_f-1)\rangle nN/2$, and the two dotted lines show
the averaged standard deviation.
The central part of
the time dependence $|\tau| < 200$ is omitted in averaging
(see text).

\item[Fig.7]
Time dependence of the segment fluctuations 
for data in Fig.1 with $n=32,\ L=30$, and $t_s=20,\,30,\,40$.
Middle curves show the first reduced moment, Eq.(4.15a);
the lower ones are for the second moment, Eq.(4.15b), and the
upper for the fluctuation parameter $F^2$, Eq.(4.15c).

\item[Fig.8]
An approximate scaling of the fluctuation law 
$P_{wr}(\tau /t_s)$
for data in Fig.1 with $n=32,\ L=30$, and 
$t_s=10,\ 20,\,30,\,40$ (symbols). The theory with empirical parameter $b=30$, Eq.(4.16), is 
presented by solid lines.

\item[Fig.9]
Two different realizations of a big fluctuation
are shown by wiggly lines,
black and gray with dots.
A smooth solid line is semiempirical relation (3.13) for the averaged big fluctuation, and
two oblique straight lines represent
diffusive kinetics near the maximum, Eq.(3.14).

\item[Fig.10]
Correlation of local fluctuations (5.1) within the window
$\tau =300 - 8492$
for data in Fig.1 with $N=4,\ n=32$ 
(thick line), and its
approximation by empirical relation (5.2) (thin line).

\end{itemize}

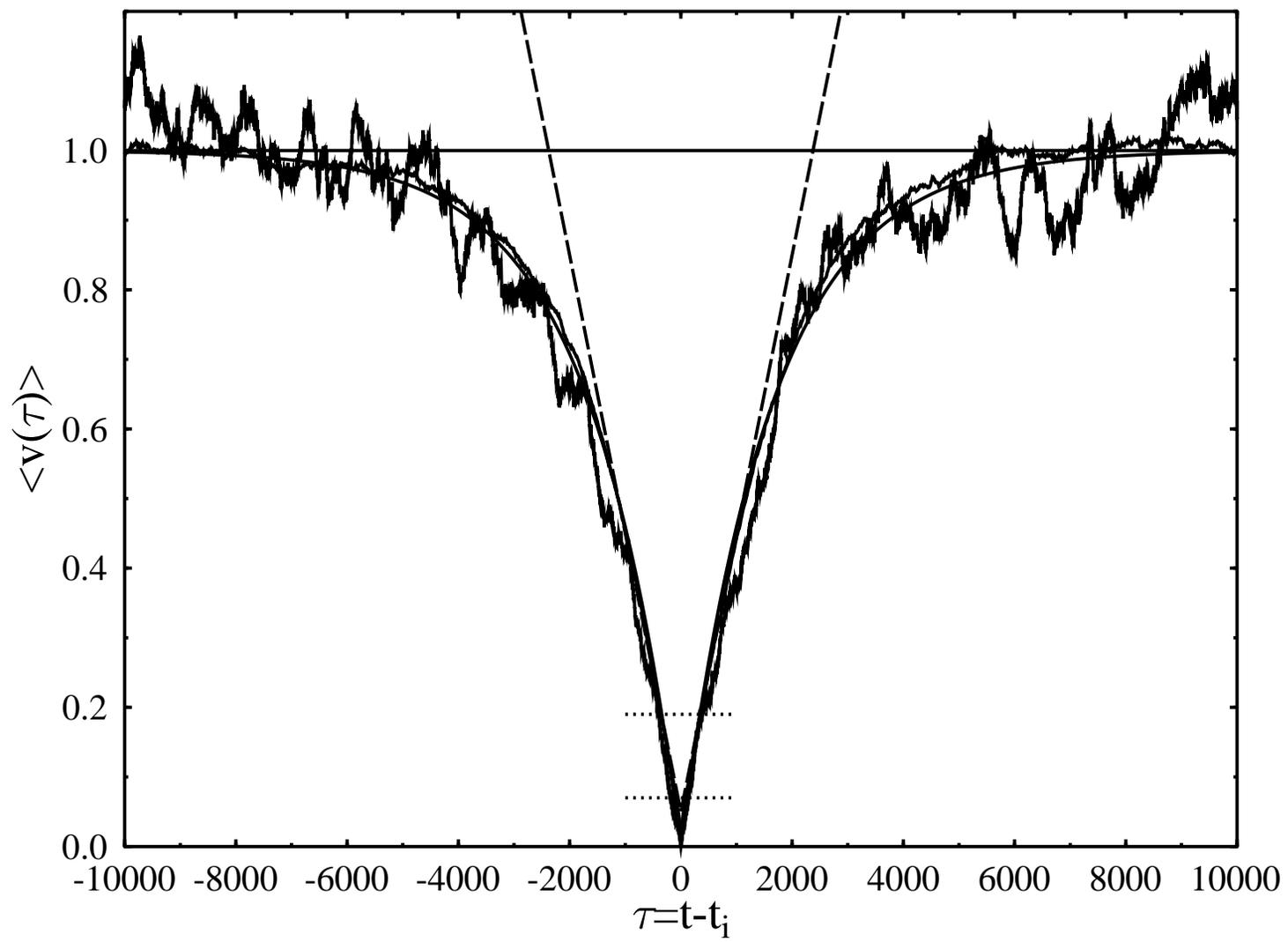
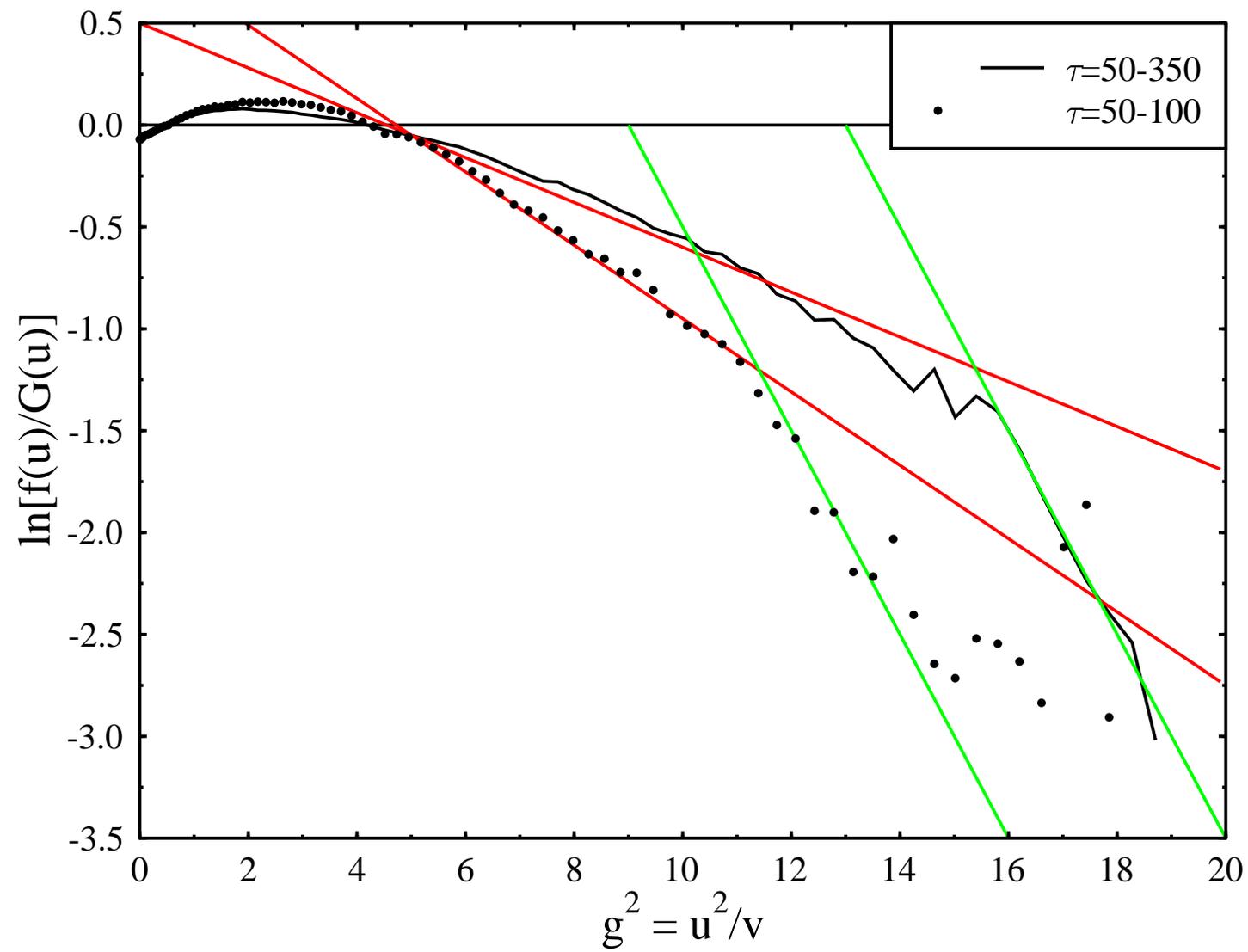
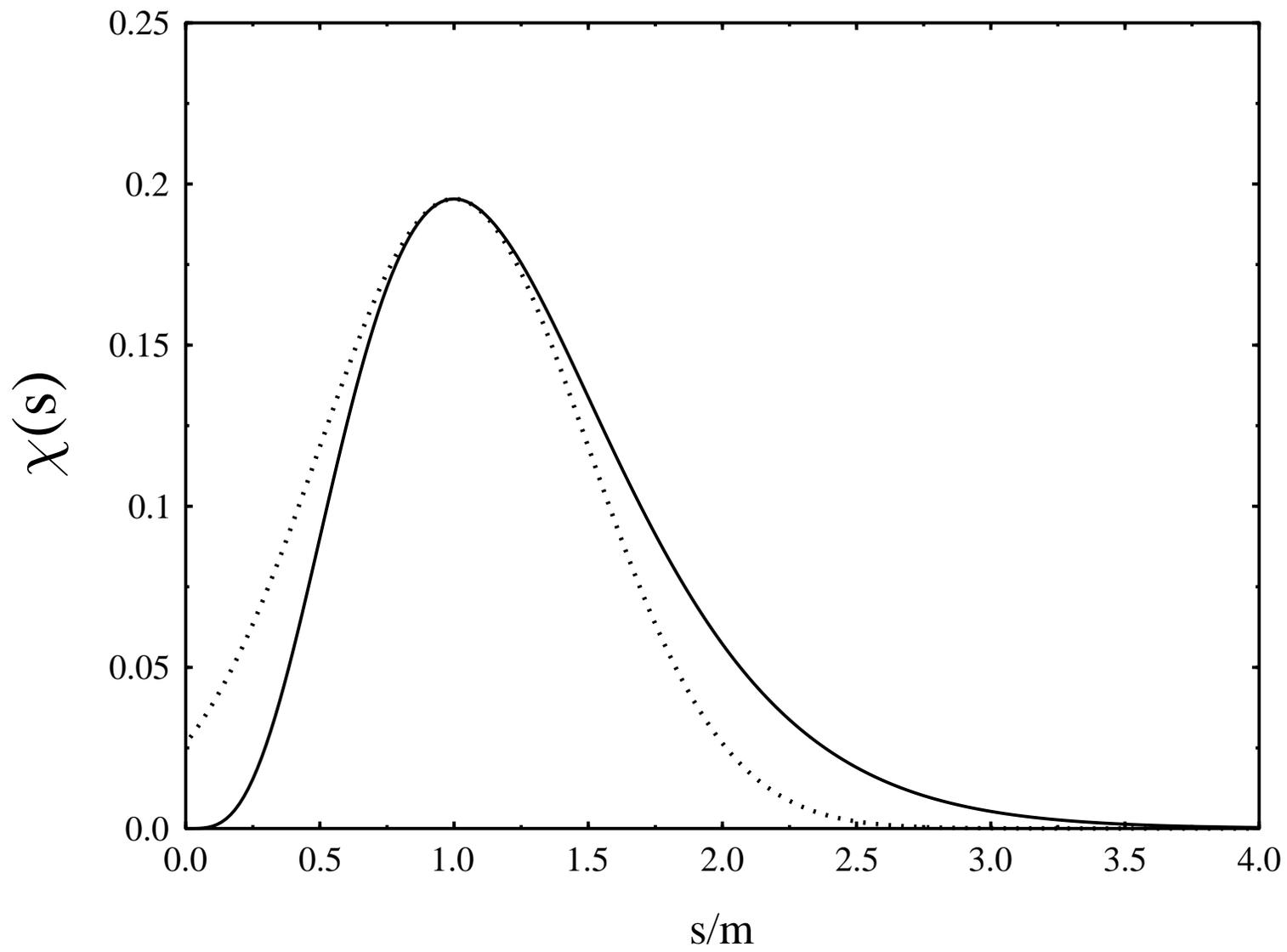
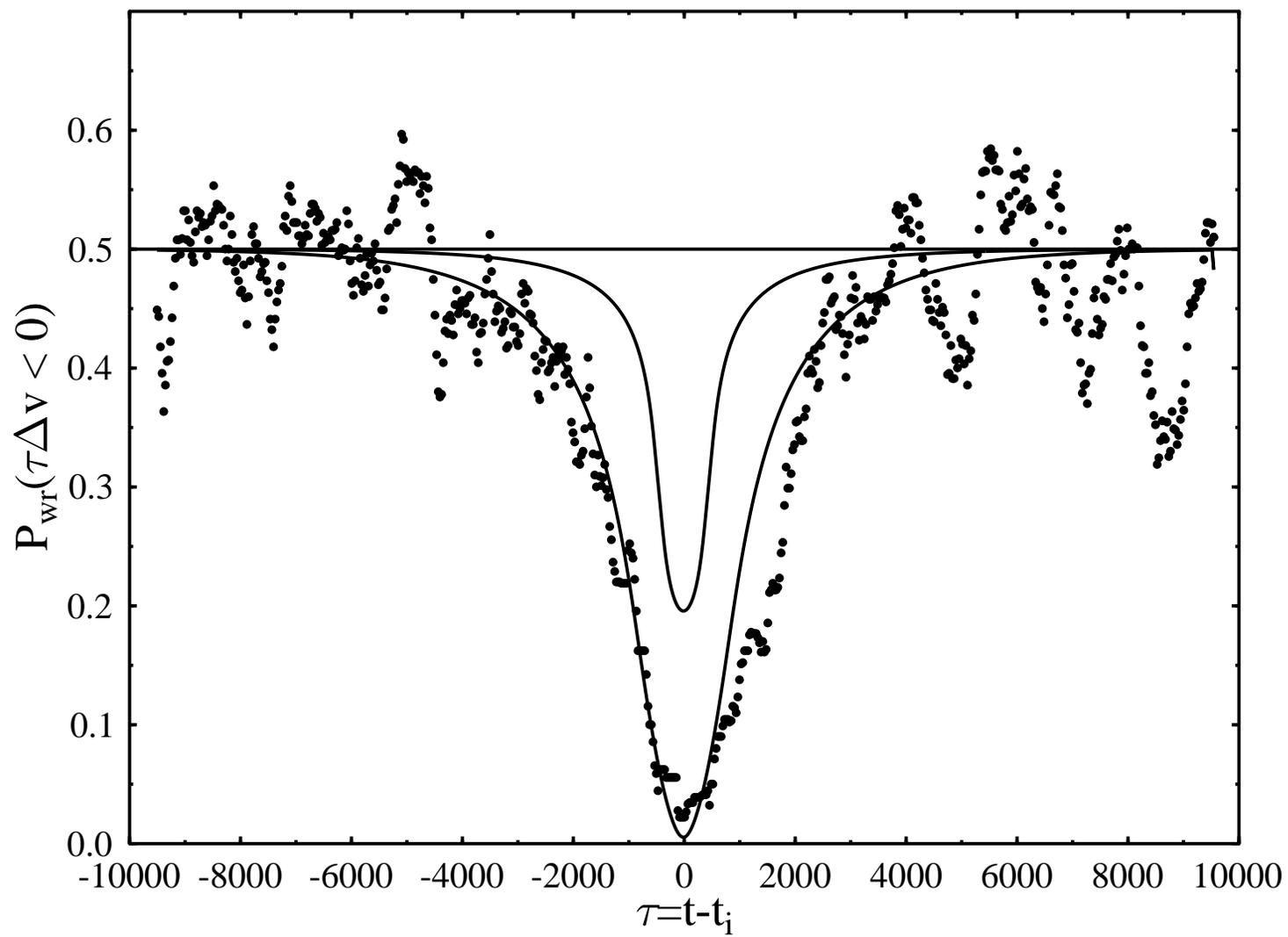
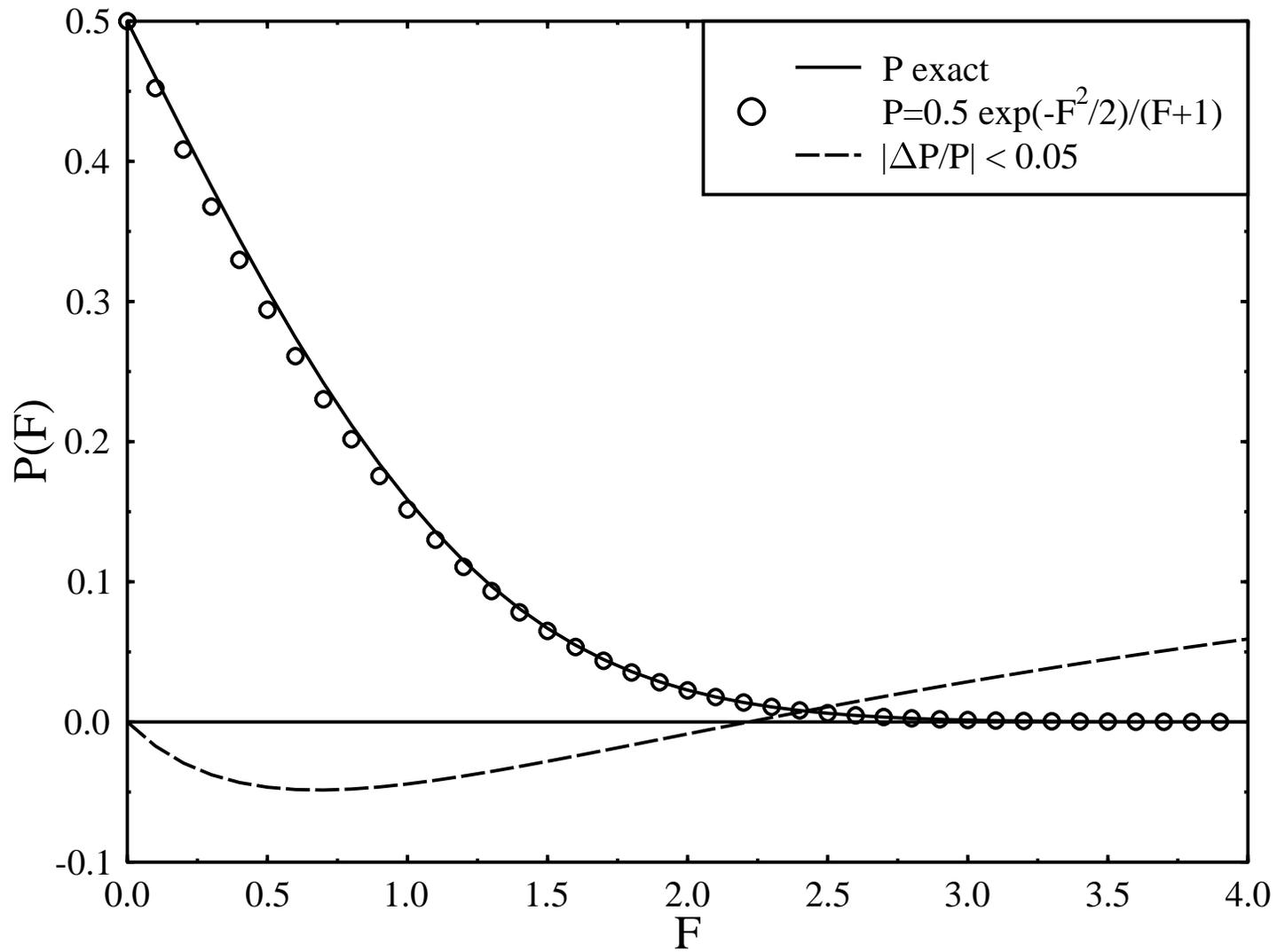
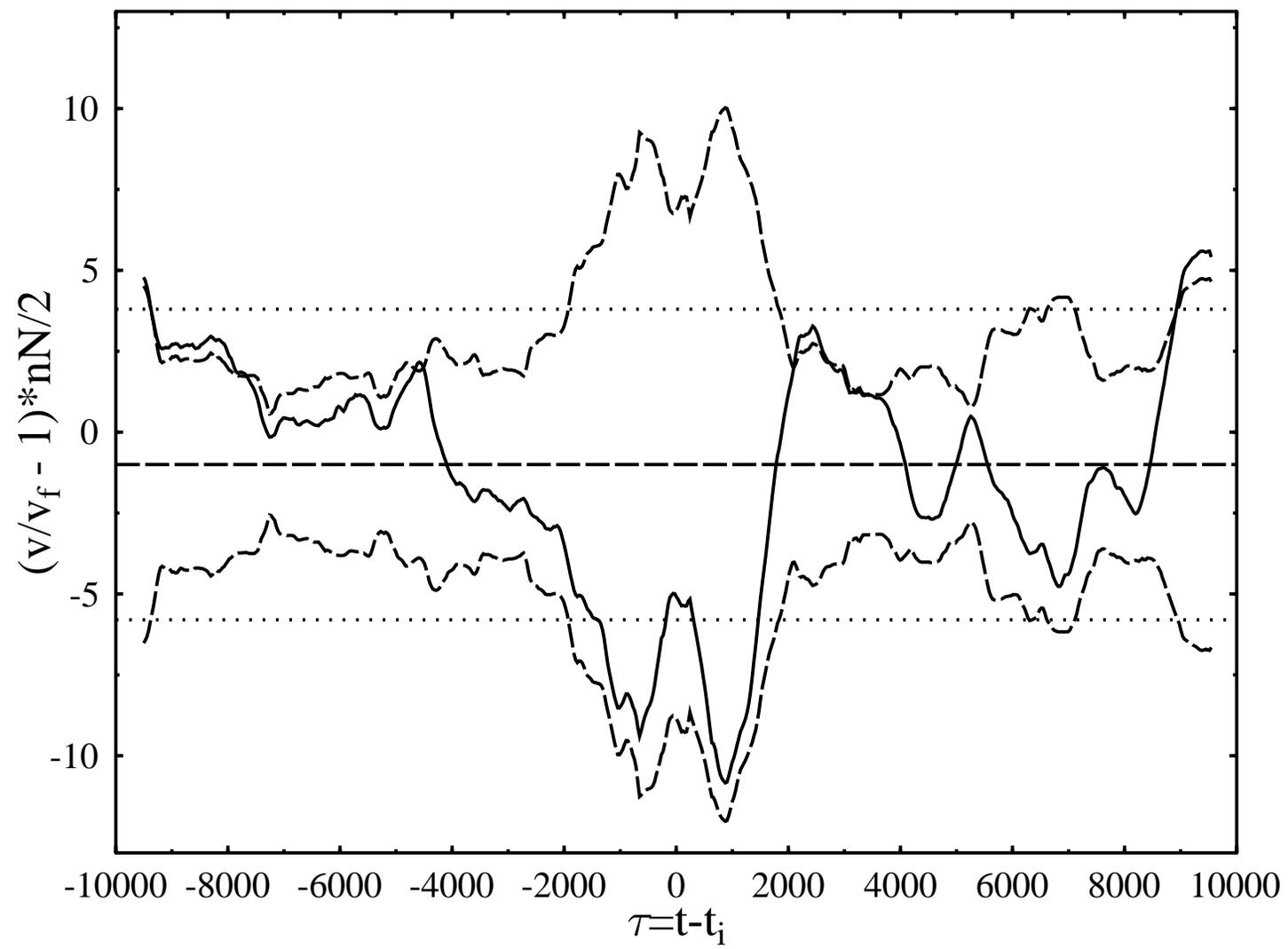
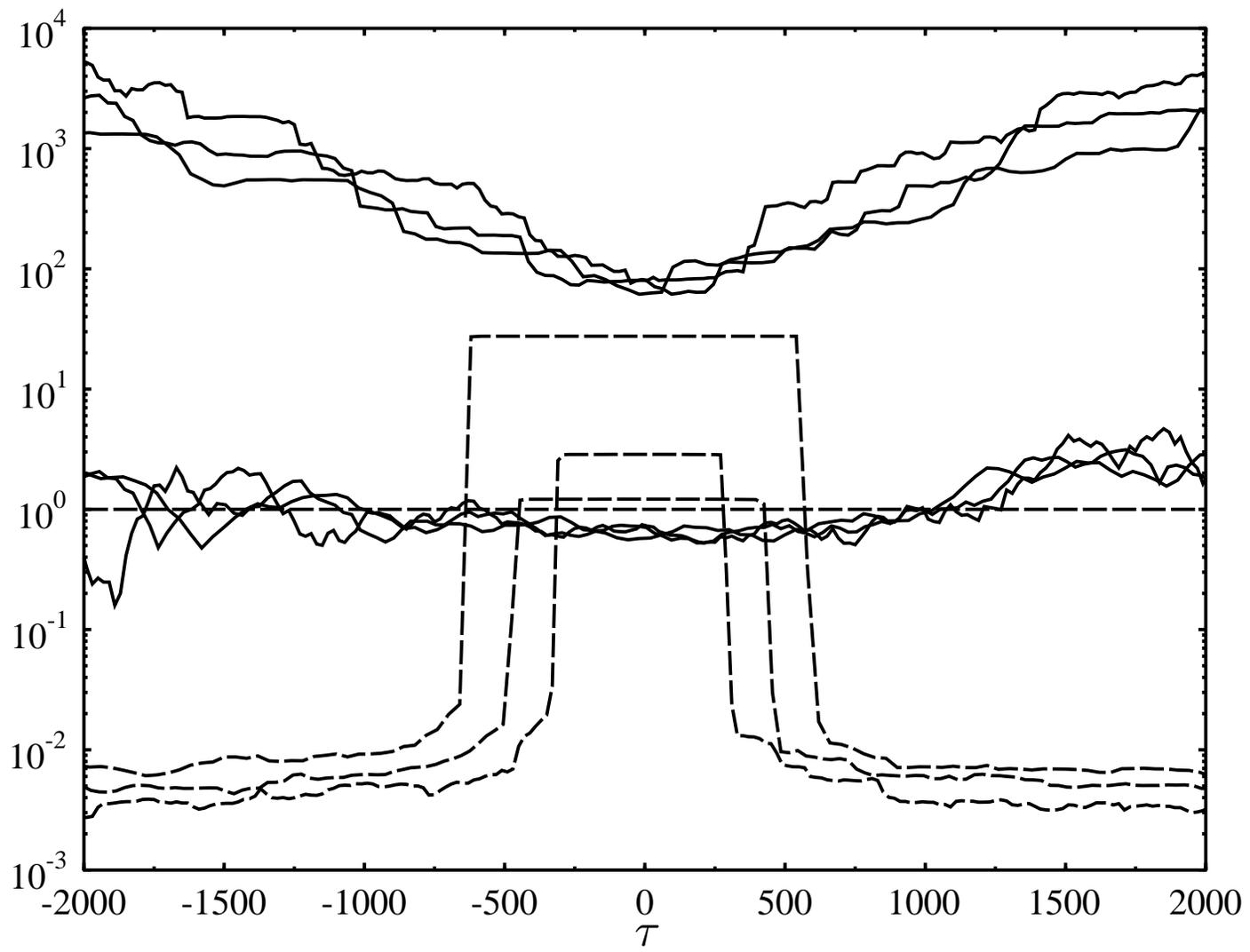
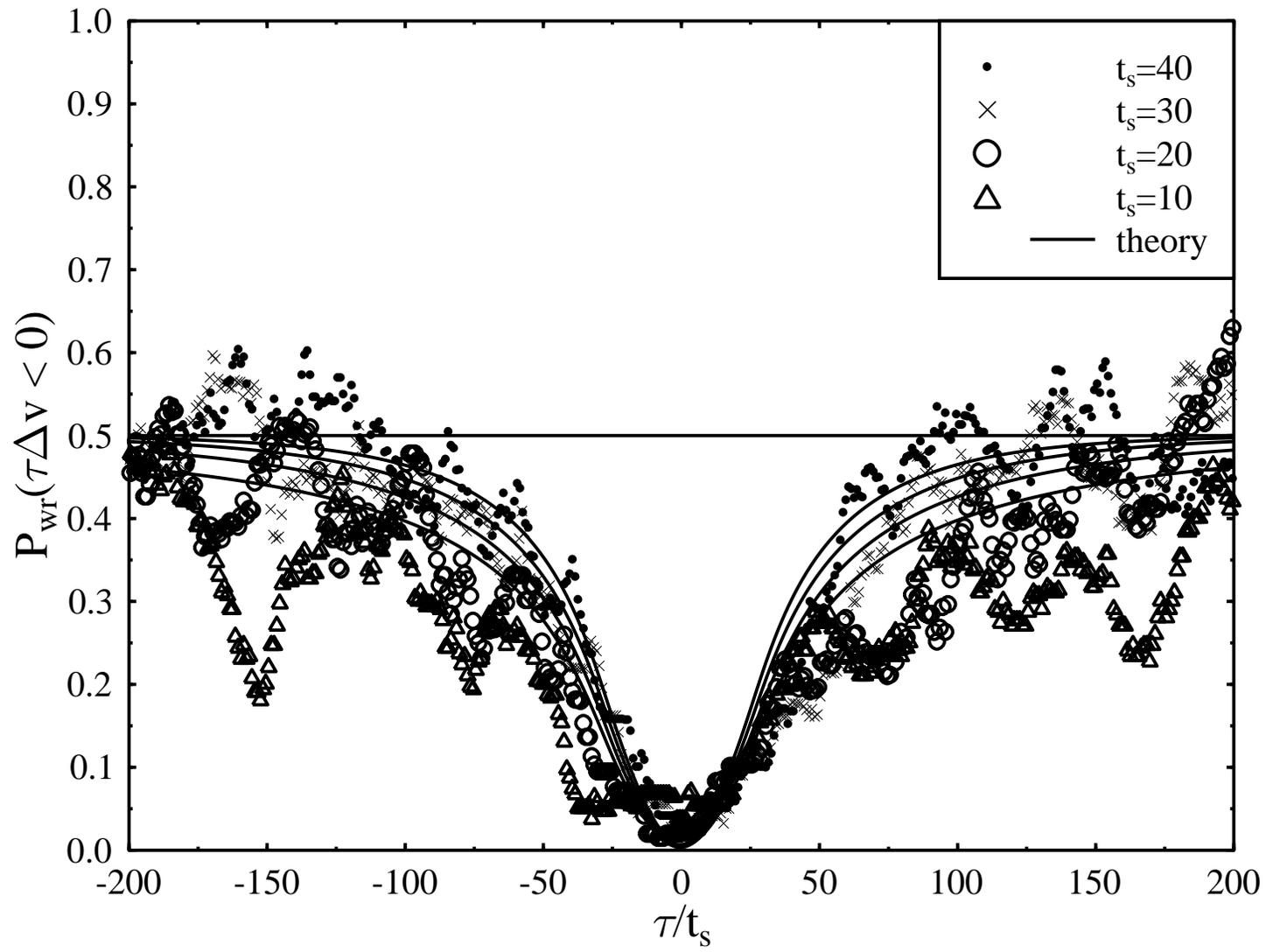
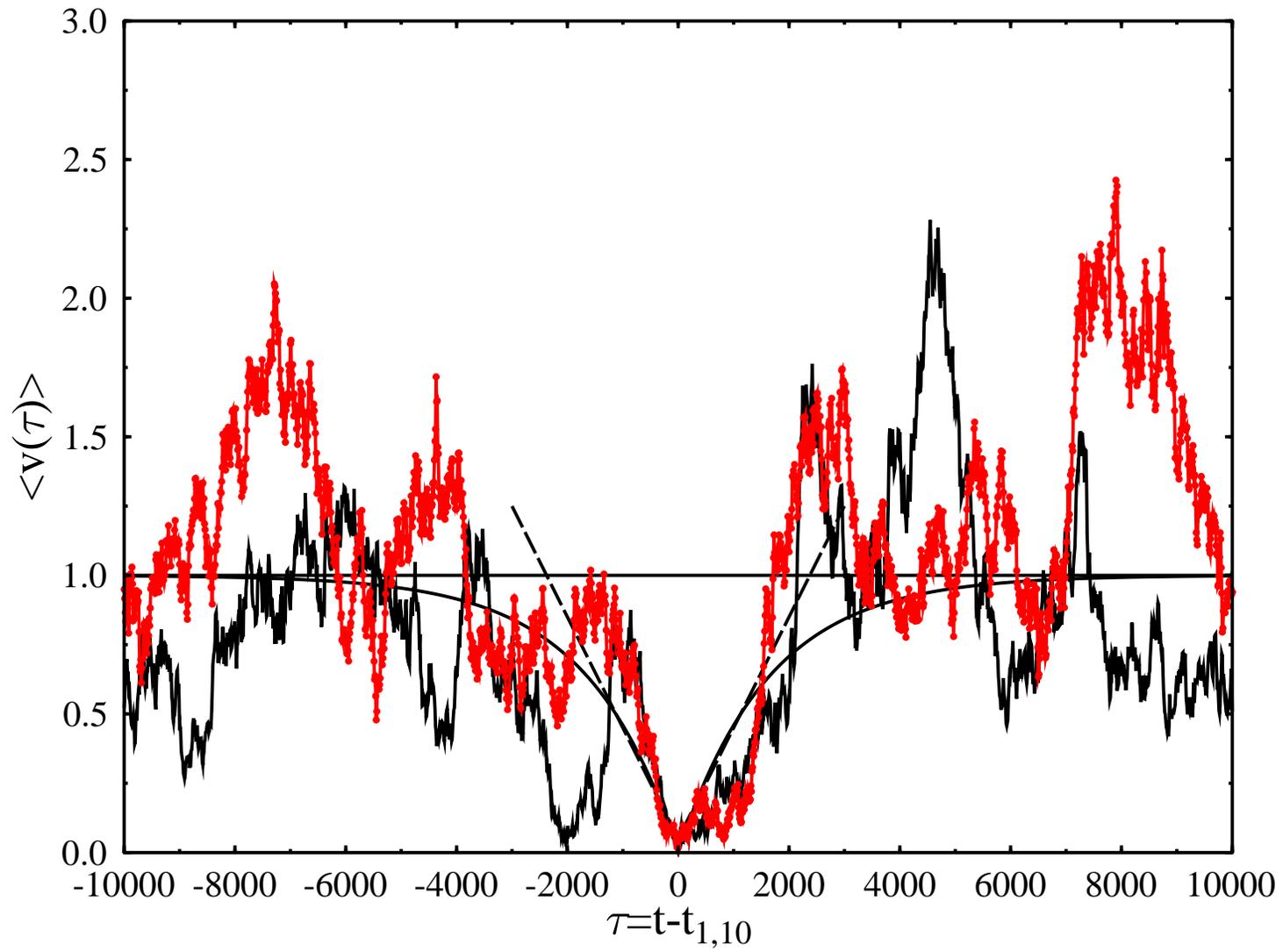
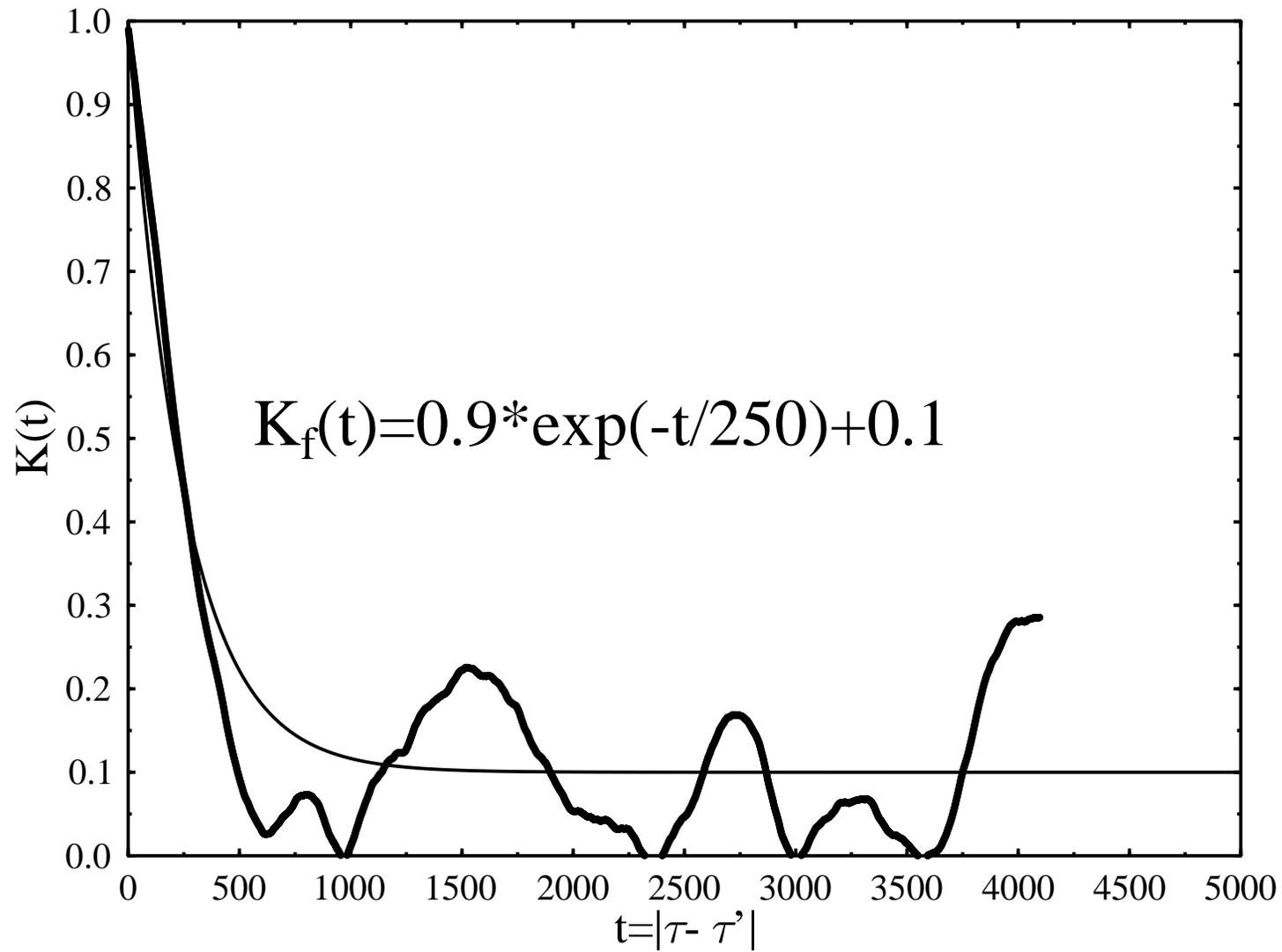
\end{document}